\documentclass[aps,prl,superscriptaddress,twocolumn,floatfix,bibnotes]{revtex4}
\usepackage{graphicx}
\usepackage[]{amsmath}

\begin{document}


\title{Control via electron count of the competition between magnetism and
superconductivity in cobalt and nickel doped NaFeAs}


\author{Dinah R. Parker}
\affiliation{Department of Chemistry, University of Oxford, Inorganic
Chemistry Laboratory, South Parks Road, Oxford, OX1 3QR, United Kingdom}
\author{Matthew J. P. Smith}
\affiliation{Department of Chemistry, University of Oxford, Inorganic
Chemistry Laboratory, South Parks Road, Oxford, OX1 3QR, United Kingdom}
\author{Tom~Lancaster}
\author{Andrew~J.~Steele}
\author{Isabel~Franke}
\affiliation{Clarendon Laboratory, University of Oxford, Parks Road,
Oxford OX1 3PU, United Kingdom}
\author{Peter~J.~Baker}
\affiliation{ISIS Facility, STFC-Rutherford Appleton Laboratory,
Harwell Science and Innovation Campus, Didcot, OX11 0QX, United
Kingdom}
\author{Francis~L.~Pratt}
\affiliation{ISIS Facility, STFC-Rutherford Appleton Laboratory,
Harwell Science and Innovation Campus, Didcot, OX11 0QX, United Kingdom}
\author{Michael~J.~Pitcher}
\affiliation{Department of Chemistry, University of Oxford, Inorganic
Chemistry Laboratory, South Parks Road, Oxford, OX1 3QR, United Kingdom}
\author{Stephen~J.~Blundell}
\email{s.blundell@physics.ox.ac.uk}
\affiliation{Clarendon Laboratory, University of Oxford, Parks Road,
Oxford OX1 3PU, United Kingdom}
\author{Simon~J.~Clarke} 
\email{simon.clarke@chem.ox.ac.uk}
\affiliation{Department of Chemistry, University of Oxford, Inorganic
Chemistry Laboratory, South Parks Road, Oxford, OX1 3QR, United Kingdom}



\date{\today}

\newcommand{\chem}[1]{\ensuremath{\mathrm{#1}}}

\begin{abstract}
Using a combination of neutron, muon and synchrotron techniques we
show how the magnetic state in NaFeAs can be tuned into
superconductivity by replacing Fe by either Co or Ni. Electron count
is the dominant factor, since Ni-doping has double the effect of
Co-doping for the same doping level.  We follow the structural,
magnetic and superconducting properties as a function of doping to
show how the superconducting state evolves, concluding that the
addition of 0.1 electrons per Fe atom is sufficient to traverse the
superconducting domain, and that magnetic order coexists with
superconductivity at doping levels less than 0.025 electrons per Fe
atom.
\end{abstract}

\maketitle


Since the discovery of high-temperature superconductivity in
fluoride-doped LaFeAsO \cite{kamihara}, several series of compounds
containing fluorite-type or anti-PbO type iron pnictide or
chalcogenide layers with a range of intervening layers have been shown
to exhibit unconventional superconductivity \cite{review_hosono}. The
superconducting pnictides range from compounds in which the iron
pnictide layers are separated by electropositive metal cations to
those in which a much thicker perovskite-type oxide slab separates the
iron pnictide layers and the electronic dimensionality of the system
is correspondingly much lower \cite{ogino}. As for the layered cuprate
superconductors, the superconducting regime is adjacent to
antiferromagnetic order.  The boundary between these regions may be
traversed by changing the electron count, by applying an external
hydrostatic pressure or by making an isoelectronic chemical
substitution which changes the structural parameters, essentially
applying a ``chemical pressure''.

\begin{figure}
\includegraphics[width=8.5cm]{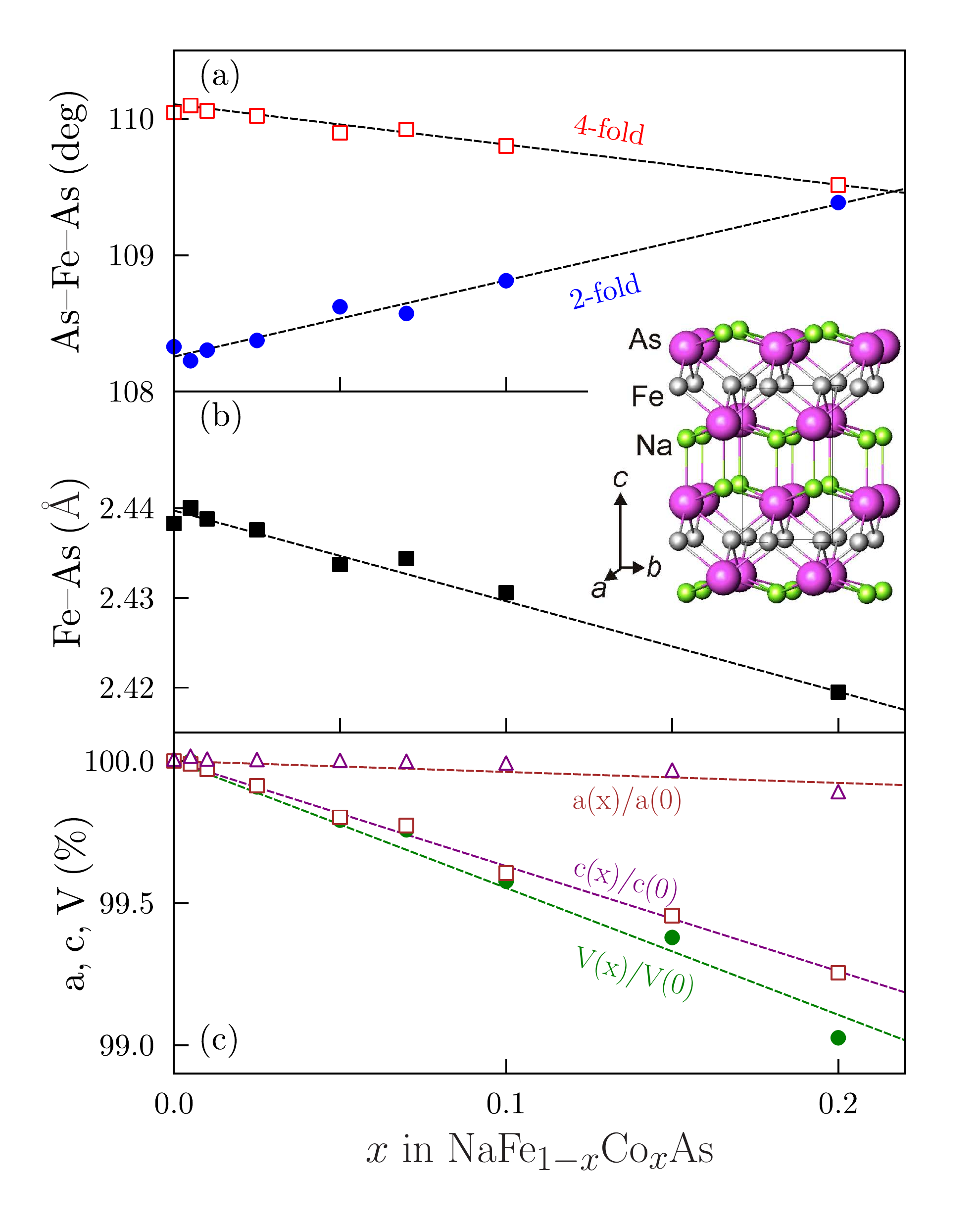}
\caption{(Color online.) The evolution at 298\,K of (a) 
As--Fe--As bond angles (b) lattice parameters
and (c) Fe--As bond lengths
of NaFe$_{1-x}$Co$_x$As as a function of Co content. The lines are
straight-line fits.  Inset: crystal structure of NaFeAs.
\label{structure}}
\end{figure}

The chemically simplest superconductors containing iron arsenide
layers have formula $A$FeAs where $A$=Li, Na and adopt the anti-PbFCl
structure type [see the inset to Fig.~\ref{structure}(b)]. LiFeAs was
shown to be a bulk superconductor in its undoped, stoichiometric form
\cite{pitcher,tapp}. Stoichiometric NaFeAs \cite{parker} also showed
evidence for superconductivity: a sample which appeared to consist of
a single crystallographic phase with very close to full occupancy of
the Na site, and which had a 10\% superconducting volume fraction
evident from DC magnetometry, also exhibited long-range
antiferromagnetic ordering in most of its volume as demonstrated by
muon-spin rotation ($\mu$SR) measurements \cite{parker}.  These
results imply that there is some overlap between the
antiferromagnetically ordered and superconducting regions of the phase
diagram.  In this Letter we present high-resolution powder
diffraction, $\mu$SR, and magnetometry measurements on
NaFe$_{1-x}$M$_x$As (M = Co, Ni) that show the extremely high
sensitivity of the competition between magnetic ordering and
superconductivity in NaFeAs derivatives to the electron count.

The compounds prepared were NaFe$_{1-x}$Co$_x$As ($x =0$, 0.005, 0.01,
0.025, 0.05, 0.07, 0.1, 0.15, 0.2) and NaFe$_{1-x}$Ni$_x$As ($x =
0.025$, 0.05, 0.1, 0.2). All are extremely air sensitive and all
manipulations of solids were performed in an argon-filled glove
box. The products were prepared from stoichiometric quantities of
elemental reagents: freshly cut sodium pieces were placed in a 9~mm
diameter Nb tube and a well-ground mixture of the transition metal and
arsenic powders were placed on top. The tube was sealed under 1\,atm
of argon in an arc-welding furnace and heated at 200$^\circ$C in a
resistance furnace for 48 hours. The black initial product was
extracted from the tube, homogenized in an agate mortar and pressed
into a pellet which was heated in a second sealed Nb tube at
750$^\circ$C for 48 hours. The tube was cooled to room temperature in
a few hours at the natural rate of the furnace. The final products
were highly crystalline lustrous black powders and appeared phase pure
according to the results of laboratory X-ray powder diffraction
measurements.

Structural investigations were carried out using the high-resolution
X-ray diffractometer ID31 at the ESRF, Grenoble, France and the
high-resolution neutron powder diffractometer HRPD at the ISIS
facility, UK. On ID31 the samples were contained in 0.7\,mm diameter
glass capillary tubes sealed closed under helium exchange gas. Samples
for neutron diffraction were contained in vanadium cylinders sealed
with indium gaskets. Rietveld refinement was performed using the GSAS
suite \cite{vondreele}.  The diffraction data did not indicate more
than 1--2\% Na deficiency, consistent with earlier reports
\cite{parker,dai}. The evolution of the lattice parameters and key
structural parameters with composition at room temperature are
summarized in Fig.~\ref{structure}.  The Fe(Co)As$_4$ tetrahedra
become more regular as Fe is substituted by Co
[Fig.~\ref{structure}(a)] and the Fe(Co)--As bond distances decrease
almost monotonically [Fig.~\ref{structure}(b)] as the smaller Co atom
is substituted.  The almost monotonic contraction in cell volume (by
$\approx 1$\% between $x = 0$ and $x = 0.2$) is dominated by the
contraction of the $c$ lattice parameter ($\approx 0.8$\% over this
range of compositions, compared to $\approx 0.1$\% for the basal
lattice parameter, $a$ [Fig.~\ref{structure}(c)]).  A similar effect
is found for Ni doping.

\begin{figure}
\includegraphics[width=8.5cm]{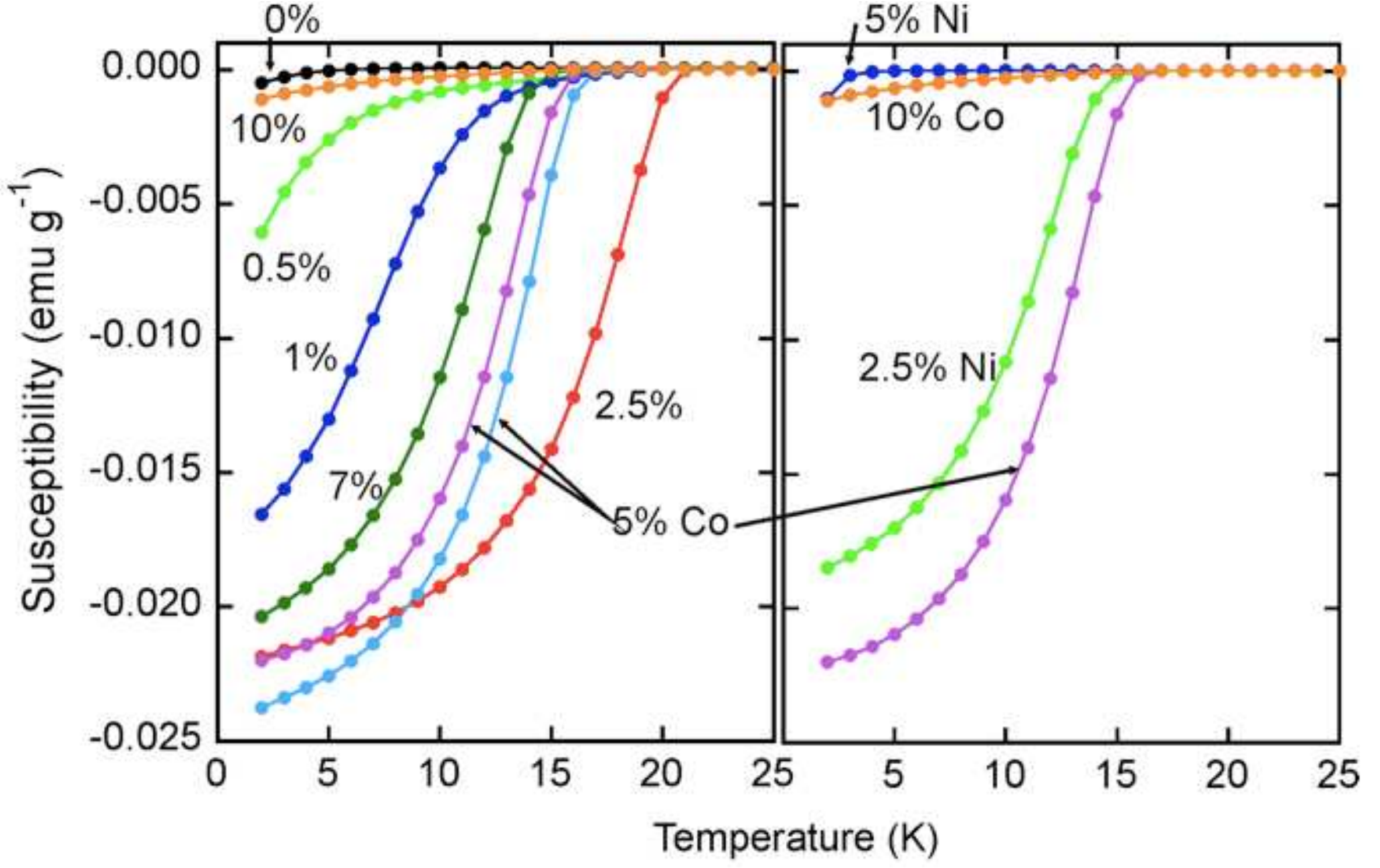}
\caption{(Color online.) The evolution of the zero-field magnetic
susceptibility (measured in an applied magnetic field of 5\,mT) of
Na$_{1-x}$Co$_x$As samples (left) and the comparison of Ni-and Co-doped
samples (right) showing that the susceptibilities are similar for
isoelectronic compounds. Doping levels are expressed as percentages
of Co unless specified.
\label{magnetometry}}
\end{figure}

Magnetometry measurements were carried out on a Quantum Design MPMS-XL
SQUID magnetometer under zero-field cooled and field cooled conditions
in a measuring field of 5\,mT.  The evolution of the low temperature
magnetic susceptibility with composition is shown in
Fig.~\ref{magnetometry}. The reported behavior of NaFeAs
\cite{parker,chen,chu} is reproducible with several samples showing
very small superconducting fractions (5--10\%) and broad
superconducting transitions with $T_{\rm c}$ of around 10 K as judged
from the onset of diamagnetism. Small levels of electron doping away
from NaFeAs produce a rapidly-increasing superconducting fraction and
in the case of Co doping, which supplies one additional electron per
Co atom, a fully superconducting state is attained between
NaFe$_{0.975}$Co$_{0.025}$As and NaFe$_{0.93}$Co$_{0.07}$As with the
maximum $T_{\rm c}$ of 21 K attained in NaFe$_{0.975}$Co$_{0.025}$As.

Substituting Ni for Co at an equivalent doping level doubles the
amount of electron doping but produces a very similar doping
dependence of lattice parameters.  As shown in
Fig.~\ref{magnetometry}, the isoelectronic compounds
NaFe$_{0.95}$Co$_{0.05}$As and NaFe$_{0.975}$Ni$_{0.025}$As exhibit
extremely similar susceptibilities.  Similarly, at the electron-rich
boundary of the superconducting region, the isoelectronic
NaFe$_{0.9}$Co$_{0.1}$As and NaFe$_{0.95}$Ni$_{0.05}$As both show very
tiny superconducting fractions.  These results strongly suggest that
electron count, and hence band filling, is the key parameter which
controls the superconductivity in this series of compounds. Samples
with higher doping levels than 0.1 electrons per Fe did not exhibit
superconductivity.

\begin{figure}
\includegraphics[width=8.5cm]{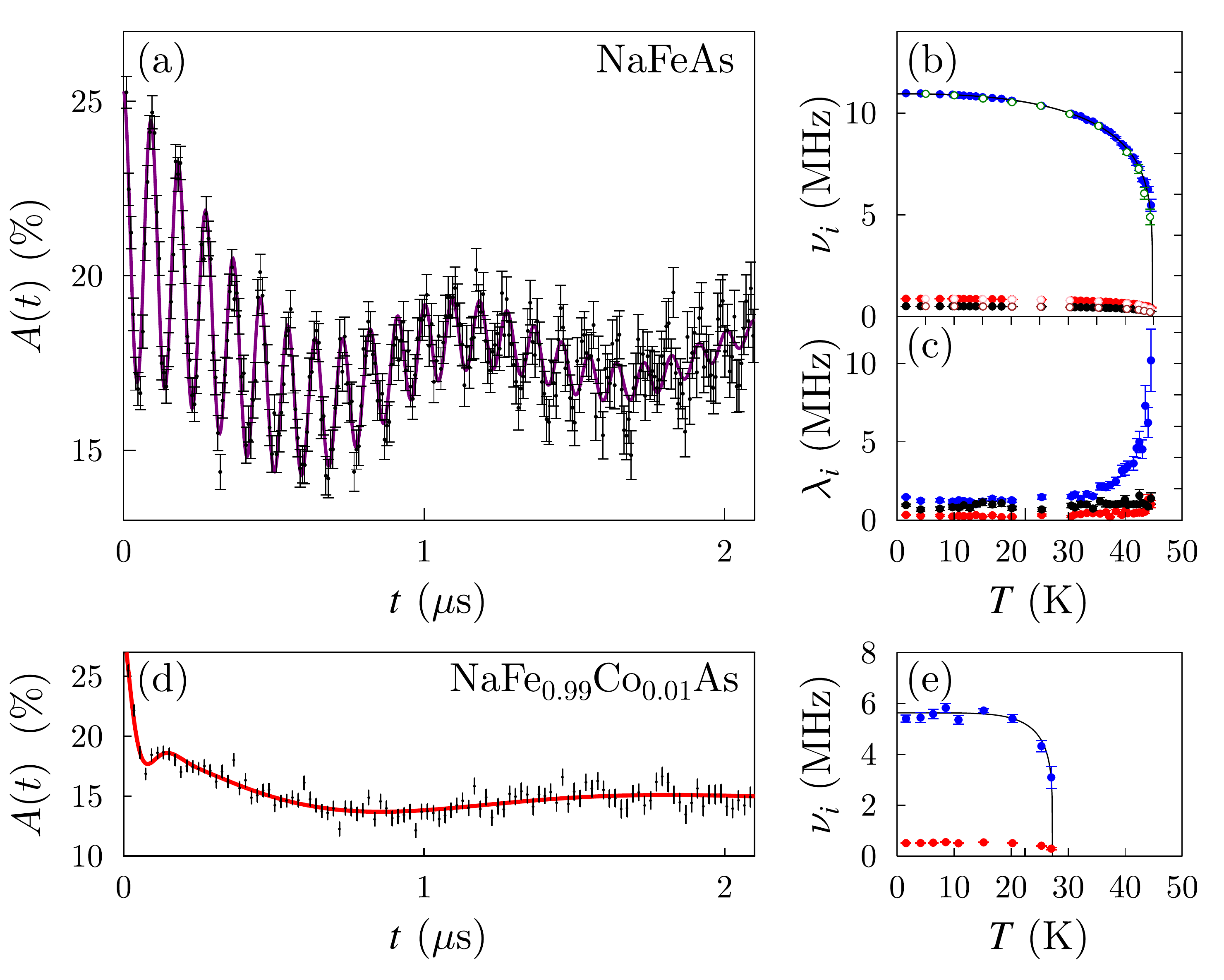}
\caption{(Color online.) (a) 
Zero-field $\mu$SR data in NaFeAs at
1.8\,K.
(b) Temperature dependence of the three precession frequencies (data
for two different samples are shown in open and closed symbols).  (c)
Relaxation rate for the three frequencies.
(d) Zero-field $\mu$SR data in NaFe$_{0.99}$Co$_{0.01}$As at
1.8\,K and
(e) temperature dependence of the  precession frequencies (only two
frequencies can be extracted in this case).
\label{muon}}
\end{figure}

Zero-field $\mu$SR experiments were carried out on the GPS beamline at
the Swiss Muon Source (S$\mu$S).  In NaFeAs, a clear oscillatory
signal is observed [Fig.~\ref{muon}(a)] which can be fitted to an
expression which is the sum of three components: $A(t)=\sum_{i=1}^3
A_i \cos(2\pi\nu_i t) {\rm e}^{-\lambda_i t}$.  All three precession
frequencies $\nu_i$ can be fitted and each follows a temperature
dependence [Fig.~\ref{muon}(b)] consistent with a N\'eel temperature
$T_{\rm N}=45.0(2)$\,K and a critical exponent $\beta=0.20(2)$; These
results are consistent with antiferromagnetic order with a primarily
two-dimensional character.  The relaxation rates $\lambda_i$ stay
roughly constant with temperature, but one of them (corresponding to
the highest frequency component) diverges at $T_{\rm N}$.  On
substituting 1\% of the Fe ions with Co to obtain a sample with a
superconducting fraction of about 60\%, as judged by magnetometry, and
a broad superconducting transition [Fig.~\ref{magnetometry}], there is
still evidence for long-range magnetic ordering in the sample, but the
oscillations in the asymmetry are heavily damped [Fig.~\ref{muon}(d)]
and lower in frequency and $T_{\rm N}$ [Fig.~\ref{muon}(e)].  This
implies a smaller internal field with a more inhomogeneous field
distribution.

\begin{figure}
\includegraphics[width=8.5cm]{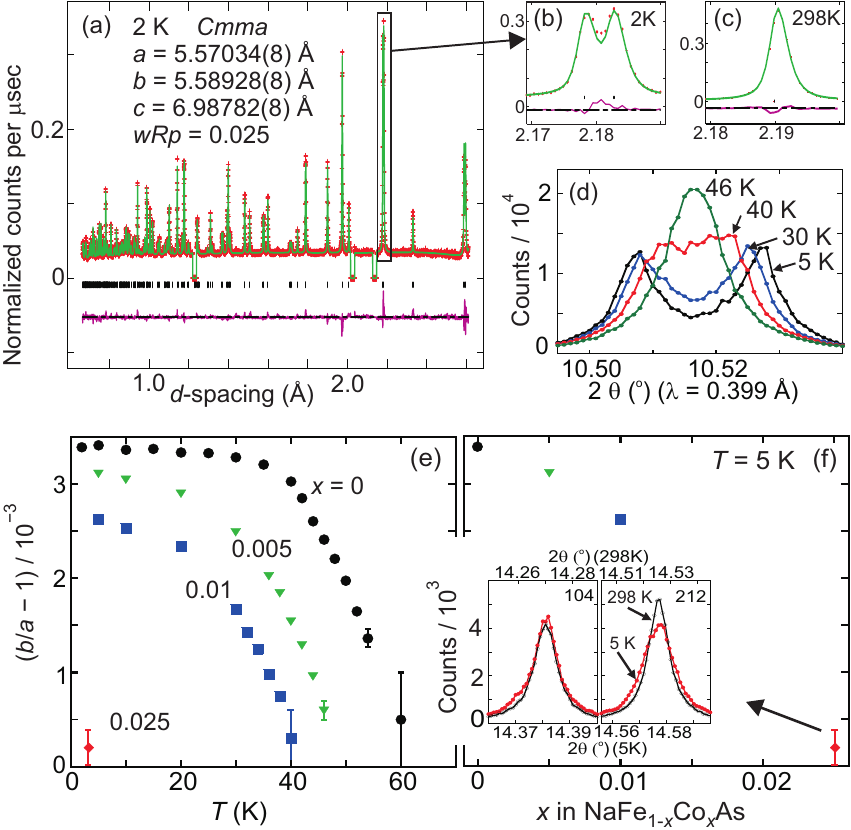}
\caption{(Color online.)  (a) Rietveld refinements of NaFeAs against
  2\,K neutron data (HRPD). Data (red, crosses), fit (green solid
  line) and difference plots are shown. Scattering from the sample
  environment has been excluded.  (b)--(c) Detail showing the
  splitting of the tetragonal 112 reflection into the orthorhombic 022
  and 202 reflections.  (d) Evolution of the orthorhombic 022/202
  reflections in NaFe$_{0.995}$Co$_{0.005}$As measured on ID31; At
  46\,K the orthorhombic and tetragonal models give fits of comparable
  quality.  (e) Temperature dependence of the distortion for
  NaFe$_{1-x}$Co$_x$As. The error bars lie within the points except
  when the presence of the distortion is marginal.  (f) The size of
  the distortion at 5\,K as a function of composition.  Inset: for
  NaFe$_{0.975}$Co$_{0.025}$As the tetragonal 212 reflection measured
  on ID31 at 5\,K is broader than at room temperature while the
  tetragonal 104 reflection (orthorhombic 114) does not broaden.
\label{hrpd}}
\end{figure}

The undoped and lightly Co-doped samples were measured using 
high-resolution diffraction measurements at low temperatures. The
non-superconducting ``parent'' phases of these arsenide
superconductors commonly show a structural distortion driven by the
long-range antiferromagnetic ordering. Such a distortion was not
evident in the original report on NaFeAs \cite{parker}, but it was
observed in a higher resolution study \cite{dai} and the symmetry of
the distortion was observed to be similar to that in the ``1111'' and
``122'' parent phases with an orthorhombic $\sqrt{2}a_{\rm T}\times
\sqrt{2}a_{\rm T}\times c_{\rm T}$ cell of $Cmma$ symmetry adopted at low
temperatures where $a_{\rm T}$ and $c_{\rm T}$ are the tetragonal lattice
parameters. On HRPD the splitting of, for example, the 112 reflection
in $P4/nmm$ into the 022 and 202 in $Cmma$ is well resolved
[Fig.~\ref{hrpd}(a)--(c)]. The high-resolution measurements show that
in NaFeAs the distortion does not vanish until 55--60\,K, well above
the antiferromagnetic N\'eel temperature similar to the behaviour
observed for all classes of parent materials apart from the
$A$Fe$_2$As$_2$ ($A$=Ca, Sr, Ba) ``122'' series where the two
transitions are coincident. Our measurements confirm that the two
transitions identified by Chen et al.\ \cite{chen} at 41\,K and 52\,K
from heat capacity measurements correspond to the
antiferromagnetic-ordering transition at 45\,K and the structural
distortion at about 55\,K.

\begin{figure}
\includegraphics[width=7.5cm]{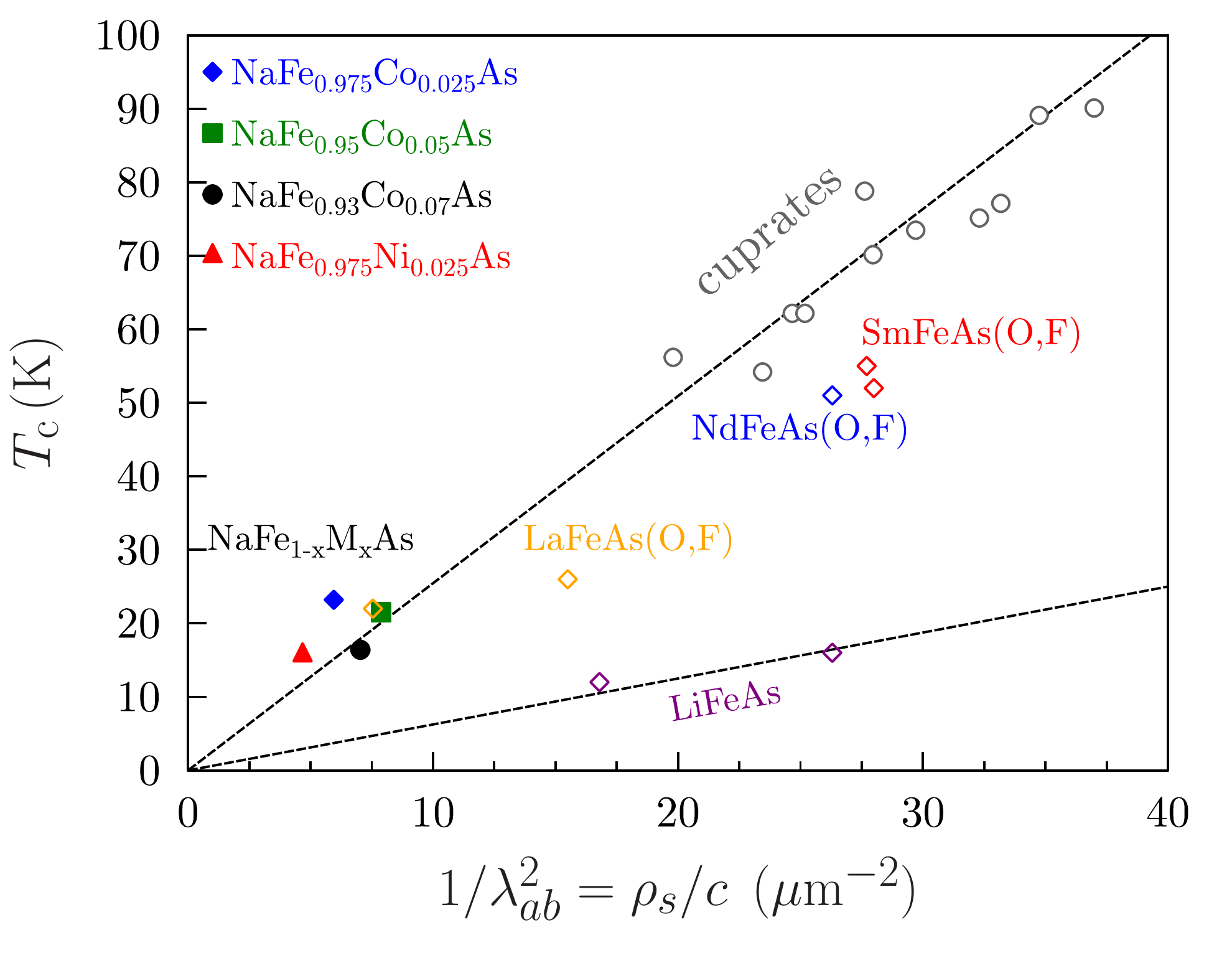}
\caption{(Color online.)  Uemura plot of the superconducting
  transition temperature $T_{\rm c}$ versus the low temperature
  superfluid stiffness. Data obtained here for the doped NaFeAs
  samples are compared with previously reported data for other iron
  arsenides and cuprates \cite{la,sm,nd,uemuramusr}.  The lower dashed
  line indicates the trend line for electron-doped cuprates, which
  closely corresponds to the behavior of the LiFeAs system \cite{flp},
  in contrast to the other pnictide superconductors, which sit closer
  to the trend for hole-doped cuprates.  The symbol size represents
  the typical estimated uncertainty of the data points.
\label{uemura}}
\end{figure}

\begin{figure}
\includegraphics[width=8.5cm]{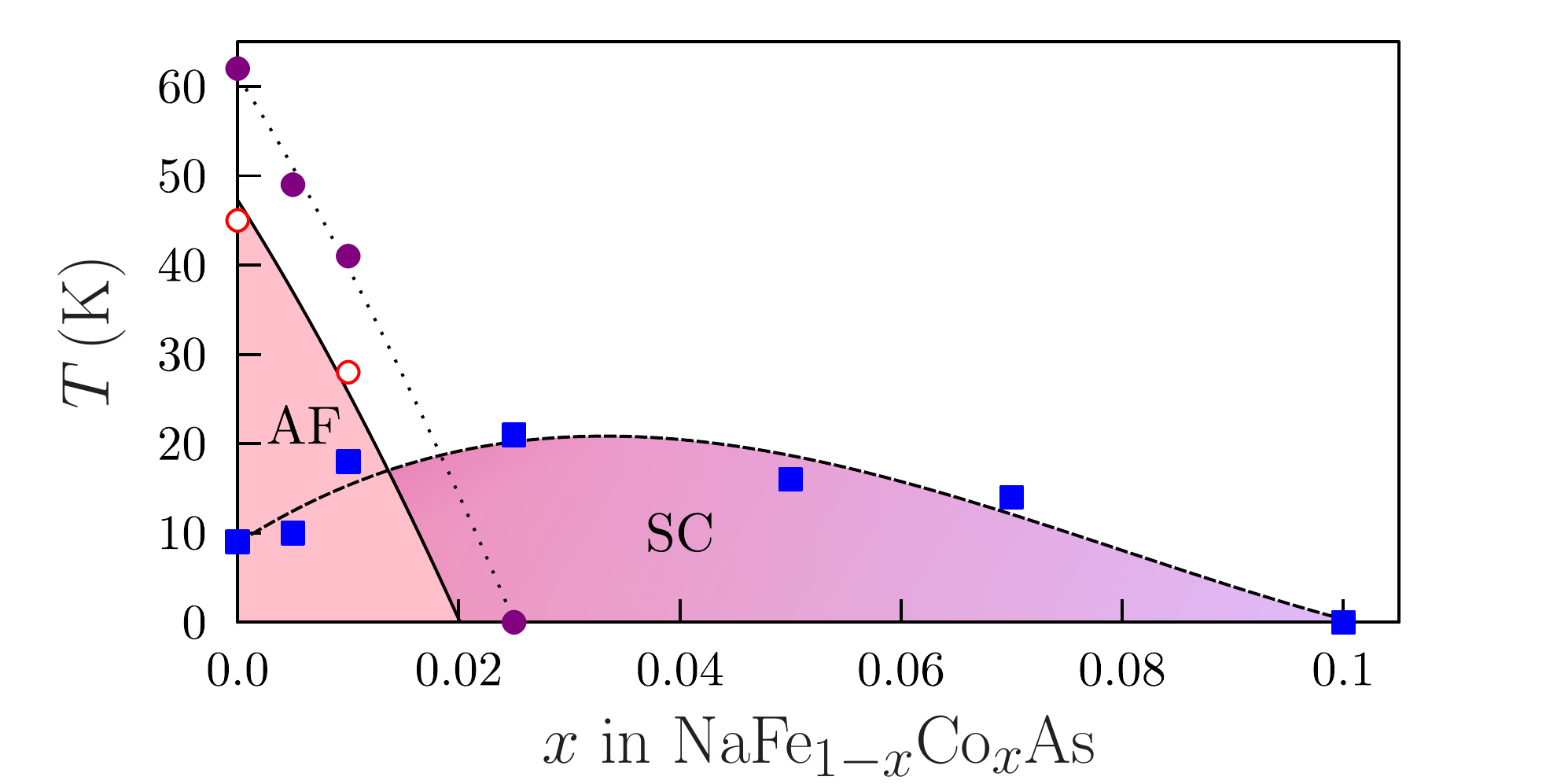}
\caption{(Color online.) Phase diagram for
NaFe$_{1-x}$Co$_x$As.  The maximum temperature at which the distortion
persists is shown by the filled circles and dotted line.
The antiferromagnetic (AF) phase is delineated by $T_{\rm N}$ (open
circles)
and superconducting (SC) phase by $T_{\rm c}$ (filled squares).
\label{phasedia}}
\end{figure}

In the lightly Co-doped regime $x\leq 0.01$ in which $\mu$SR shows
that antiferromagnetic ordering persists, diffraction measurements on
the high-resolution powder diffractometer ID31 show that the
distortion ($b_{\rm O}/a_{\rm O}-1$, in terms of the basal plane
orthorhombic lattice parameters) persists above $T_{\rm N}$
[Fig.~\ref{hrpd}(d)--(e)].  The size of the distortion at 5\,K
decreases smoothly with increasing Co content [Fig.~\ref{hrpd}(f)]
showing a similar dependence on composition to the N\'eel temperature.

In iron pnictide systems the structural distortion associated with
antiferromagnetic order is known to be quenched in the fully
superconducting regime. Measurements at 5\,K on ID31 of
NaFe$_{0.975}$Co$_{0.025}$As, which lies in this regime, revealed
broadening of the reflections which would split under an orthorhombic
distortion, while reflections which would not split retained their
ambient temperature widths [Fig.~\ref{hrpd}(f)~inset].  At 5\,K, the
orthorhombic model refined stably but the goodness of fit was
negligibly superior to that of the tetragonal model and the lattice
parameters were clearly correlated with the peak profile
parameters. An orthorhombic distortion may thus persist into the
superconducting regime, although the upper bound on $b_{\rm O}/a_{\rm
  O}-1$ is $3.0(1)\times 10^{-4}$ in NaFe$_{0.975}$Co$_{0.025}$As, an
order of magnitude smaller than in NaFeAs, and marginal even using an
extremely high-resolution diffractometer.

Once a large superconducting fraction is established ($x\geq 0.025$)
it is possible to use transverse-field $\mu$SR to study the superfluid
stiffness by measuring the temperature and field-dependence of the rms
field broadening $B_{\rm rms}$ due to the vortex lattice
\cite{sonier}.  Such experiments were performed using both S$\mu$S and
the ISIS Pulsed Muon Facility and these data can be used to deduce the
in-plane penetration depth $\lambda_{ab}$ and superfluid stiffness
$\rho_s$ using $\rho_s/c=\lambda_{ab}^{-2} = ( 3/0.00371 )^{1/2}
B_{\rm rms}/\Phi_0$.  Plotting our data on an Uemura plot
(Fig.~\ref{uemura}) demonstrates that the values of superfluid
stiffness correlate well with the trend observed in hole-doped
cuprates and other iron arsenides, in contrast to the enhanced
superfluid stiffness found for LiFeAs \cite{flp}.

The conclusions of our work are that NaFeAs lies very close to one
edge of a superconducting dome which is traversed by the addition of
0.1 electrons per Fe atom using either Co or Ni doping (see
Fig.~\ref{phasedia}).  As the system passes from the magnetically
ordered regime to the superconducting regime there is a region of
phase coexistence which has a width of less than 0.025 electrons per
Fe, so the system is extremely sensitive to composition.  The width
and shape of the superconducting dome and the width of the coexistence
region are similar to those determined in the ``122'' system
Ba(Fe$_{1-x}$Co$_x$)$_2$As$_2$ \cite{fisher} and in the ``1111''
system CaFe$_{1-x}$Co$_x$AsF \cite{matsuishi}.

We thank the EPSRC (UK) for financial support and STFC (UK) for access
to ISIS and ESRF, K. S. Knight for assistance on HRPD, and A. N. Fitch
and I. Margiolaki for assistance on ID31.  Part of this work was
performed at the Swiss Muon Source (S$\mu$S).


\begin{thebibliography}{11}
\bibitem{kamihara}
Y. Kamihara, T. Watanabe, M. Hirano, and H. Hosono,
J. Am. Chem. Soc. {\bf 130}, 3296, (2008).
\bibitem{review_hosono}
K. Ishida, Y. Nakai and H. Hosono J. Phys. Soc. Jpn, {\bf 78}, 062001,
(2009).
\bibitem{ogino} H. Ogino {\sl et al.}
Supercond. Sci. Technol. {\bf 22},  075008,  (2009).
\bibitem{pitcher}
M. J. Pitcher {\sl et al.}
Chem. Commun. 5918, (2008).
\bibitem{tapp}
J. H. Tapp {\sl et al.}
Phys. Rev. B {\bf 78}, 060505(R) (2008). 
\bibitem{parker}
D. R. Parker {\sl et al.}
Chem. Commun. 2189 (2009).
\bibitem{vondreele} A. Larson and R. B. von Dreele, The General Structure Analysis
System, Los Alamos National Laboratory, Los Alamos, NM,
(1985).
\bibitem{dai}
S. Li {\sl et al.}
Phys. Rev. B {\bf 80}, 020504(R) (2009).
\bibitem{chen}
G. F. Chen, W. Z. Hu, J. L. Luo, and N. L. Wang,
Phys. Rev. Lett. {\bf 102}, 227004, (2009).   
\bibitem{chu} 
C. W. Chu
{\sl et al.}
Physica C {\bf 469}, 326, (2009).
\bibitem{sonier}
J. E. Sonier, J. H. Brewer, and R. F. Kiefl, Rev. Mod. Phys. {\bf 72},
769 (2000).
\bibitem{la}
H. Luetkens
{\sl et al.}
Phys. Rev. Lett. {\bf 101}, 097009 (2008).
\bibitem{sm}
A. J. Drew
{\sl et al.}
Phys. Rev. Lett. {\bf 101}, 097010 (2008);
R. Khasanov
{\sl et al.}
Phys. Rev. B {\bf 78}, 092506 (2008).
\bibitem{nd}
J. P. Carlo
{\sl et al.}
Phys. Rev. Lett. {\bf 102}, 087001 (2009).
\bibitem{uemuramusr}
Y. J. Uemura, 
Physica B {\bf 374}, 1 (2006).
\bibitem{flp}
F. L. Pratt
{\sl et al.}
Phys. Rev. B {\bf 79},  052508,  (2009).
\bibitem{fisher} 
J.-H. Chu, J. G. Analytis, C. Kucharczyk and I. R. Fisher, Phys. Rev. B {\bf 79}, 014506, (2009).
\bibitem{matsuishi}
S. Matsuishi
{\sl et al.}
J. Am. Chem. Soc. {\bf 130}, 14428, (2008).


\end{thebibliography}
\end{document}